\documentclass[12pt]{article}
\usepackage{epsfig}
\usepackage{amsfonts}
\usepackage{amsopn}
\usepackage{amsmath}
\usepackage{verbatim,amsthm}

\DeclareMathOperator{\ad}{ad}
\DeclareMathOperator{\diag}{diag}
\DeclareMathOperator{\op}{op}
\DeclareMathOperator{\re}{Re}
\DeclareMathOperator{\Spin}{\textrm{\emph{Spin}}}
\DeclareMathOperator{\supp}{supp}
\DeclareMathOperator{\tr}{tr}
\DeclareMathAlphabet{\mathpzc}{OT1}{pzc}{m}{it}
\newtheorem{thm}{Theorem}
\newtheorem*{main_thm}{Main theorem}
\newtheorem{lem}[thm]{Lemma}

\newtheorem{prop}[thm]{Proposition}

\title{Octonionic twists for supermembrane\\matrix models}

\author{
Jens Hoppe$^a$ \thanks{e-mail:  hoppe@math.kth.se} \ ,  \
Douglas Lundholm$^a$  \thanks{e-mail: dogge@math.kth.se} \\
and Maciej Trzetrzelewski$^{a,b}$ \thanks{e-mail: 33lewski@th.if.uj.edu.pl}  \\
\\
$^a$  Department of Mathematics,\\
Royal Institute of Technology, \\
KTH, 100 44 Stockholm, \\
Sweden \\ \\
$^b$ Institute of Physics,\\
Jagiellonian University, \\
Reymonta 4, 30-059 Krak\'ow,\\
Poland
}

\date{}

\begin{document}

\maketitle

\begin{abstract}
A certain $G_2 \times U(1)$ invariant Hamiltonian arising from the
standard membrane matrix model via conjugating any of the
supercharges by a cubic, octonionic, exponential is proven to have a
spectrum covering the whole half-axis $\mathbb{R}_+$. The model
could be useful in determining a normalizable zero-energy state in
the original $SO(9)$ invariant $SU(N)$ matrix model.
\end{abstract}

\setlength\arraycolsep{2pt}
\def\arraystretch{1.5}

\section{Introduction}

Despite considerable effort \cite{1} during the last decade, and crucial relevance to
M-theory \cite{Witten-BFSS} / membrane theory \cite{GH,BST,dWHN} /
reduced Yang-Mills theory \cite{bjorken,dimred,baake}, existence, uniqueness and
structure of zero-energy states in $\Spin(9)\times SU(N)$ invariant
supersymmetric matrix models are not really understood to a degree that one could call satisfactory.

In this paper we consider models with  $G_2 \times U(1) \times SU(N)$ 
symmetry that we obtain by deforming (cp. \cite{PR,EHS})
the $\Spin(9) \times SU(N)$ models, and which we believe to be relevant
both from the point of view of deformation theory and possible
relations between ground states, as well as because (for the fixed
value of the deformation parameter that we take) the Hamiltonian is
slightly simpler, and
therefore a good testing ground for new approaches.

The model is introduced in Section 2 by deforming the $\Spin(9)\times
SU(N)$ invariant one via a particular cubic exponential. In Section 3, with
the help of various propositions that are proved in Section 4,
this model is shown to share a central feature of the original theory,
namely that the Hamiltonian, 
in contrast with the disctreteness of the spectrum (cp. \cite{Simon-Luscher}) for the purely bosonic theory,
has an essential spectrum covering the whole positive axis (cp. \cite{dWLN}).
A summary of the results is presented in Section 5.
In the appendices some background material is provided,
and the deformation we introduce put into a slightly more general context.

\section{The deformed model}

To find a normalizable state annihilated by the
supercharges \footnote{cp. Appendix A}
$$
	Q_{\beta}:=\left(i\delta_{\alpha
	\beta}\frac{i}{2}f_{ABC}z_B\bar{z}_C +  i\Gamma_{\alpha \beta}^j
	\frac{\partial}{\partial
	x_{jA}}-\frac{1}{2}f_{ABC}x_{jB}x_{kC}\Gamma^{jk}_{\alpha\beta}
	\right)\lambda_{\alpha A}
$$
\begin{equation}
	+ \left(2\delta_{\alpha\beta}\frac{\partial}{\partial z_A}
	-if_{ABC}x_{jB}\bar{z}_C \Gamma^j_{\alpha\beta}
	\right)\lambda_{\alpha A}^{\dagger}  \label{charge}
\end{equation}
(and by their hermitian conjugates) is a difficult task; the
$(x_{jA})_{j=1,\ldots,7}^{A=1,\ldots,N^2-1}$ and $z_A = x_{8A} + ix_{9A}$ are bosonic coordinates,
$(\lambda_{\alpha A})_{\alpha=1,\ldots,8}^{A=1,\ldots,N^2-1}$
Grassmann variables, $f_{ABC}$ totally antisymmetric structure
constants of $SU(N)$, 
$\Gamma^{jk} := \frac{1}{2}[\Gamma^j,\Gamma^k]$,
and $(\Gamma^j)_{j=1,\ldots,7}$ (purely imaginary, antisymmetric) matrices
satisfying $\{\Gamma^j,\Gamma^k\}=2\delta^{jk}\mathbf{1}_{8 \times 8}$, in a
particular representation given by $i\Gamma^j_{\alpha 8}=\delta^j_{\alpha}$,
$i\Gamma^j_{kl}=-c_{jkl}$, totally anti-symmetric octonionic
structure constants.

In \cite{Hoppe2} conjugation by the exponent of
\begin{equation}
	g(x):=\frac{1}{6}f_{ABC}x_{jA}x_{kB}x_{lC} \left(i\Gamma^{jkl} \right)_{\beta \beta}
\end{equation}
was shown to remove the third term in \eqref{charge} (extending an
observation made in \cite{dWHN}; note that $\sum_j \Gamma^j_{\alpha
\beta}\Gamma^{jkl}_{\beta \beta}=\Gamma^{kl}_{\alpha \beta}$ for fixed $\beta$, cp. Appendix B).
Correspondingly, defining $H_k := \{Q(k),{Q(k)}^{\dagger}\} \ge 0$,
where (choosing $\beta=8$)
\begin{equation}
	Q(k) :=e^{k g(x)}Q_8 e^{-k g(x)}= Q_8 +
	\frac{k}{2}f_{ABC}x_{jB}x_{lC} \Gamma^{jl}_{\alpha 8}
	\lambda_{\alpha A},
\end{equation}
gives
\begin{eqnarray}
    H_k &=&
    -\Delta_{{\mathbb{R}}^{9(N^2-1)}} + (k-1)^2V_{1\ldots 7} + V_{89}
    + \bar{z}_A f_{ABC}x_{jB} f_{A'B'C}x_{jB'} z_{A'} \nonumber \\[5pt]
&&  +\ 2f_{AA'E}x_{jE}( \delta_{\alpha 8}\delta_{\alpha' j} - \delta_{\alpha j}\delta_{\alpha' 8} )\lambda_{\alpha A}\lambda_{\alpha' A'}^{\dagger} \nonumber \\[5pt]
&&  +\ 2(k-1)f_{EAA'}x_{jE}\left( i\Gamma^j \right)_{ll'} \lambda_{l A}\lambda_{l'A'}^{\dagger} \nonumber \\[5pt]
&&  +\ f_{EAA'} z_E \lambda_{\alpha A}\lambda_{\alpha A'}
    + f_{EAA'} \bar{z}_E \lambda_{\alpha A'}^{\dagger} \lambda_{\alpha A}^{\dagger}. \label{H_k}
\end{eqnarray}
The potential terms for the $x$- resp. $z$-coordinates are given by
$$
    V_{1\ldots 7} = \frac{1}{2} f_{ABC}x_{jB}x_{lC} f_{AB'C'}x_{jB'}x_{lC'}
    \quad \textrm{resp.} \quad
    V_{89} = \frac{1}{4} f_{ABC}\bar{z}_{B}z_C f_{AB'C'}z_{B'}\bar{z}_{C'}.
$$

While for large $k$,
\begin{equation}
    \hat{H} := -\Delta_x + V_{1\ldots 7} - 2f_{EAA'}x_{jE}c_{jll'}\lambda_{lA} \lambda^{\dagger}_{l'A'},
\end{equation}
appears to be the relevant operator (having rescaled $x \to
(k-1)^{-1/3}x$) \footnote{This point (and \cite{EHS} in general) was
discussed with B. Durhuus and J. P. Solovej, -which we gratefully
acknowledge.}
 we will, in this note, exclusively study $H_{k=1}=:\tilde{H}$, which is of the form (cp. (\ref{H_k}))
\begin{eqnarray}
    \tilde{H} &=& -\Delta_z + \bar{z}fxfxz + 2fx(\delta\delta)\lambda \lambda^{\dagger} \nonumber \\[5pt]
    && -\Delta_x +V_{89}+ fz\lambda \lambda +f \bar{z} \lambda^{\dagger}\lambda^{\dagger}. \label{htilde}
\end{eqnarray}

The first line (arising from the second line of \eqref{charge} alone) is
denoted by $H_D \ (=H_D(x)\ge 0)$ and we will heavily use that its
spectrum and eigenfunctions are known \cite{Hoppe3}.
The operator $V_{89}+zf\lambda\lambda + \bar{z}f
\lambda^{\dagger} \lambda^{\dagger}$ appearing in the second line
will be denoted by $K$.
We also note that, regardless of the choice of $k$, 
the bosonic part of $H_k$ (first line in \eqref{H_k}) 
has a strictly positive and purely discrete spectrum 
(this is easily proved along the lines of \cite{Simon-Luscher}).

The bosonic part of $H_D$ describes two sets of $n := N^2-1$ harmonic
oscillators whose frequencies $\omega_A$ are the square root of the
eigenvalues of the parametrically $x$-dependent, positive semidefinite frequency matrix
\begin{equation}
    S_{AA'}(x):=f_{ABC}x_{jB}f_{A'B'C}x_{jB'},
\end{equation}
while its fermionic part,
$2W_{\alpha A\thinspace\beta B}(x)\lambda_{\alpha A}\lambda_{\beta B}^{\dagger}$, that is linear in $x_{jA}$,
has eigenvalues arising from those of $2W_{\alpha A\thinspace\beta B}$, 
which are $\{\pm 2\omega_A(x)\}_{A=1,\ldots,n}$ as well as $6n$
times the eigenvalue zero -- altogether leading to the exact zero-energy state(s) \cite{Hoppe3}
\begin{equation}
    \psi_x= \prod_{A=1}^{n} \sqrt{\frac{\omega_A(x)}{\pi}} \thinspace
    e^{-\frac{1}{2}\omega_A(x)\bar{z}_A'z_A'} \thinspace
    e^{(-\omega_1)}_{\alpha_1A_1}(x) \ldots e^{(-\omega_l)}_{\alpha_lA_l}(x)
    \lambda_{\alpha_1A_1} \ldots \lambda_{\alpha_lA_l}, \label{psix}
\end{equation}
(where, $n \le l \le 7n$, $\omega_{l>n}=0$, and we have diagonalized S via
$z_A'=R_{AB}(x)z_{B}$, $S=R^T [\omega_A^2] R$ ),
that involves the eigenvectors $e^{(\omega)}(x)$ of the matrix $W(x)$
corresponding to eigenvalue $\omega$.
Excited states of $H_D$ are obtained by acting with
the bosonic creation operators (i.e. multiplying $\psi_x$ by the
corresponding Hermite polynomials) and/or adding fermions
corresponding to positive eigenvalues $2\omega$ (i.e. multiplying
$\psi_x$ by $e^{(+\omega)}_{\alpha A} \lambda_{\alpha A})$.

The matrix $S$ can also be written as
$$
    S(x) = -\sum_{j=1}^7 \ad_{X_j} \circ \ad_{X_j} = -[X_j,[X_j, \ \cdot \ ]],
$$
acting on the space $i\mathfrak{su}(N)$  of traceless, hermitian matrices,
with $X_j=x_{jA}T_A$, $[T_A,T_B]=if_{ABC}T_C$.
In particular, its lowest eigenvalue $\omega_{\min}^2$ is given by
$$
    \omega_{\min}^2 = \min_{e \in S^{n-1}} e_A S_{AB} e_B = \min_{\|E\| = 1} \sum_j \| [X_j,E] \|^2,
$$
where $\| \cdot \|$ here denotes the corresponding norm
on $i\mathfrak{su}(N) \cong \mathbb{R}^n$.
For $N>2$, $S(x)$ will have zero-modes not only when all matrices $X_j$
commute, but (of qualitative significance) for the
\emph{larger} space of configurations where all the $X_j$ are simultaneously
block-diagonalizable.

\section{Continuity of the spectrum of $\tilde{H}$}

In this section we formulate and prove the main theorem of the paper.
We will make use of three propositions and one lemma (which are proved in Section \ref{sec_proofs}
in order not to break the flow of the text).

\begin{main_thm}
    For any $\lambda \ge 0$ there exists a sequence $(\Psi_t)$ of
    rapidly decaying smooth $SU(N)$-invariant functions such that
    $\left\| \Psi_t \right\| = 1$, and
    \begin{displaymath}
        \left\|(\tilde{H}-\lambda)\Psi_t \right\| \to 0 \quad \textrm{as} \quad t \to \infty.
    \end{displaymath}
\end{main_thm}

\noindent In other words, the spectrum of $\tilde{H}$ (even when
restricted to the physical Hilbert space) covers the whole positive
real line, just as it is the case for the original $H_{k=0}$.
However, because of the terms that vanish for $H_{k=1}$, together
with the convenient structure of the remaining terms noted in the
previous section, we are able to construct such a sequence
explicitly without resorting to the gauge fixing procedure used in
\cite{dWLN}.

In the following, we write $\tilde{H}$ as
\begin{equation} \label{htilde_K}
    \tilde{H} = -\Delta_x + H_D(x) + K,
\end{equation}
where \cite{Hoppe3}
\begin{equation} \label{HD_short}
    H_D(x) = -4\partial_{\bar{z}} \cdot \partial_{z} + \bar{z} \cdot S(x) z + 2 W(x) \lambda \lambda^\dagger
\end{equation}
and
\begin{equation} \label{K_def}
    K(z) = \frac{1}{4} f\bar{z}zfz\bar{z} 
    + fz\lambda\lambda + f\bar{z} \lambda^\dagger \lambda^\dagger.
\end{equation}
We also point out that, since $\tilde{H}$ is an unbounded operator,
it is considered to be defined as a differential operator on the
Schwartz class $\mathcal{S}$ of smooth functions of rapid decay,
and then extends by closure or Friedrichs extension
to a self-adjoint operator in $\mathcal{H} = L^2(\mathbb{R}^{9n}) \otimes \mathcal{F}$.

Our candidate for the sequence $\Psi_t$ will be wavefunctions given
by the minimal fermion number ground state $\psi_x$ of $H_D(x)$
multiplied by some gauge invariant cut-off function $\chi_t$.
Formally, it is convenient to write the Hilbert space $\mathcal{H}$
as a constant fiber direct integral (see \cite{R-S}) over the
$x$-coordinates,
$$
    \mathcal{H} = \int_{\mathbb{R}^{7n}}^\oplus \mathpzc{h} \ dx,
$$
where
(writing $dx = d^{7n}x$, $dz = d^nx_8 d^nx_9$ for the integration measures)
$$
    \mathpzc{h} := L^2(\mathbb{R}^{2n}) \otimes \mathcal{F} = \int_{\mathbb{R}^{2n}}^\oplus \mathcal{F} \ dz
$$
is the $z$-coordinate Hilbert space
on which the operator $H_D(x) + K$ acts in each point $x$.
Hence, for any $\Psi(x,z) = \chi(x) \psi_x(z) \in \mathcal{H}$, we have
\begin{equation} \label{norm_decomposition}
    \| \Psi \|^2_\mathcal{H}
    = \int_{\mathbb{R}^{7n}} |\chi(x)|^2 \|\psi_x\|^2_\mathpzc{h} \ dx
    = \int_{\mathbb{R}^{7n}} |\chi(x)|^2 \int_{\mathbb{R}^{2n}} \|\psi_x(z)\|^2_\mathcal{F} \ dz \ dx.
\end{equation}
We also write the ground state \eqref{psix} of $H_D(x)$ in a more compact notation,
\begin{equation} \label{psix_compact}
    \psi_x(z) = \pi^{-\frac{n}{2}} s(x)^{\frac{1}{4}} e^{-\frac{1}{2}\bar{z}\cdot S(x)^{1/2} z} \xi_x,
\end{equation}
where $s := \det S$, and $\xi_x \in \mathcal{F}_n$ (i.e. $n$ fermions)
is the normalized fermionic eigenvector satisfying
\begin{equation} \label{w_eigenvector}
    W(x) \lambda \lambda^\dagger \ \xi_x = -\sum_A \omega_A\ \xi_x = - \tr \left(S(x)^{\frac{1}{2}}\right) \xi_x.
\end{equation}
We note the following:

\begin{prop} \label{prop_smooth_inv}
    $\psi_x$ is smooth (also in $x$), rapidly decaying, and $SU(N)$-invariant.
\end{prop}

\begin{prop} \label{prop_z_estimates}
    $\left\| \psi_x \right\|_\mathpzc{h} = 1$, \quad and \quad
    $\left\| |z|^k \psi_x \right\|_\mathpzc{h} \le \frac{C_k}{\omega_{\min}^{k/2}(x)}$,
    for $k=1,2,4$ and some positive constants $C_k$.
\end{prop}

\noindent
Hence, by choosing an appropriate cut-off function $\chi_t$ for the
$x$-coordinates such that $\omega_{\min}(x)\to \infty$ as $t \to \infty$,
we can make the terms in $K(z)$ arbitrarily small.
The following proposition shows that such a choice is indeed possible.

\begin{prop} \label{prop_chi}
    For any $\lambda \ge 0$ and $t$ sufficiently large there exist $SU(N)$-invariant
    cut-off functions $\chi_t \in C_0^{\infty}(\mathbb{R}^{7n})$ such that $\forall x \in \supp \chi_t$
    \begin{equation} \label{prop_chi_support}
        \omega_{\min}(x) \ge c_1 t, \qquad  c_2 t \le |x| \le c_3 t,
    \end{equation}
    and, as $t \to \infty$,
    \begin{equation} \label{prop_chi_norms}
        \left\| \chi_t \right\|_{\mathbb{R}^{7n}}=1, \quad
        \left\| \partial_{jA}\chi_t \right\|_{\mathbb{R}^{7n}} \le c_4, \quad
        \left\| (-\Delta_{\mathbb{R}^{7n}}-\lambda)\chi_t \right\|_{\mathbb{R}^{7n}} \to  0,
    \end{equation}
    where (here, and in the following) $c_{k=1,2,3,\ldots}$ are some positive constants.
\end{prop}

As a final preparation before proving the main theorem, we state
the following lemma which ensures that also certain derivatives
tend to zero.

\begin{lem} \label{lem_derivatives}
    $\left\| \partial _{jA}\psi_x \right\|_\mathpzc{h} \le \frac{c_5}{\omega_{\min}(x)}$ \ \ and \ \
    $\left\| \partial^2 _{jA}\psi_x \right\|_\mathpzc{h} \le \frac{c_6}{\omega_{\min}^2(x)}$ \ \ on \ $\supp \chi_t$.
\end{lem}

\subsection{Proof of the main theorem}

Motivated by the expression \eqref{htilde_K} and the above 
preparations, we define
$$
    \Psi_t(x,z) := \chi_t(x)\psi_x(z),
$$
where $\chi_t \in C^{\infty}_0(\mathbb{R}^{7n})$ is chosen
according to Proposition \ref{prop_chi}. 
We note that $\Psi_t$ is in the domain of $\tilde{H}$ and
by \eqref{norm_decomposition} has $\|\Psi_t\| = 1$.
Acting with $\tilde{H}$ on $\Psi_t(x,z)$, we obtain
\begin{equation} \label{hpsi}
    \tilde{H}\Psi_t =
    - \psi_x\Delta_x\chi_t - 2\sum_{j,A}\partial_{jA}\chi_t \partial_{jA}\psi_x
    - \sum_{j,A}\chi_t\partial_{jA}^2\psi_x + \chi_t K(z) \psi_x,
\end{equation}
(where we used the fact that $H_D(x) \psi_x=0$).
Subtracting $\lambda \Psi_t$ from this equation and
using Propositions \ref{prop_z_estimates}, \ref{prop_chi} and Lemma \ref{lem_derivatives}
(and that any operator on $\mathcal{F}$ is bounded)
to estimate the norms of the terms on the r.h.s. as $t \to \infty$, we find
$$
    \left\| \psi_x(-\Delta_x-\lambda)\chi_t \right\|^2 =
    \left\| (-\Delta_x-\lambda)\chi_t \right\|^2_{\mathbb{R}^{7n}} \to 0,
$$
$$
    \left\| \partial_{jA}\chi_t \thinspace \partial_{jA}\psi_x \right\|^2 \le
    \int |\partial_{jA}\chi_t|^2 \left( \frac{c_5}{\omega_{\min}} \right)^2 dx \le
    \frac{c_7}{t^2} \left\|\partial_{jA}\chi_t \right\|^2_{\mathbb{R}^{7n}} \le
    \frac{c_8}{t^2} \to 0,
$$
$$
    \left\| \chi_t\partial_{jA}^2 \psi_x \right\|^2 \le
    \int |\chi_t|^2 \left( \frac{c_6}{\omega^2_{\min}} \right)^2 dx \le
    \frac{c_9}{t^4} \left\| \chi_t \right\|^2_{\mathbb{R}^{7n}} \to 0,
$$
$$
\begin{array}{rl}
    \left\| \chi_t K(z) \psi_x  \right\|^2
    \le & \displaystyle \int |\chi_t|^2 \left( c_{10}\left\| |z|^4\psi_x \right\|_\mathpzc{h} + c_{11} \left\| |z|\psi_x \right\|_\mathpzc{h} \right)^2 dx \\
    \le & \displaystyle \int \left(\frac{c_{10}C_4}{\omega_{\min}^2} + \frac{c_{11}C_1}{\omega_{\min}^{1/2}}\right)^2 |\chi_t|^2 dx
    \le \frac{c_{12}}{t} \left\| \chi_t \right\|^2_{\mathbb{R}^{7n}} \to 0.
\end{array}
$$
Hence, $\left\|(\tilde{H}-\lambda)\Psi_t\right\| \to 0$ as $t \to \infty$.
\qed

\section{Proofs} \label{sec_proofs}

Here we present detailed proofs of the propositions and
lemma that were stated in the previous section.

\subsection{Proof of Proposition 1}

It is obvious from \eqref{psix_compact} that
$\psi_x \in \mathcal{S}(\mathbb{R}^{2n}) \otimes \mathcal{F}_n =: \mathcal{S}_n$.
Smoothness in $x$ for the scalar (bosonic) part of $\psi_x$ follows
from our requirement that $\omega_{\min}(x) > 0$, i.e. $s > 0$
for every $x$ we consider.
As for the fermionic part $\xi_x$, smoothness follows by considering
$\mathcal{F}_n$ as a real space of dimension $\binom{8n}{n}$ and, for each point $x$,
viewing $\xi_x$ as the (up to sign) unique normalized eigenvector of the
linear map $\xi \mapsto W(x)\lambda\lambda^\dagger \xi$ with eigenvalue
$-\sum_A \omega_A(x)$.
(A consistent choice of sign can be made because
we will only be working on orientable subsets of $\mathbb{R}^{7n}$.)
Smoothness of $\xi_x$ now follows from smoothness of $W(x)$
and the implicit function theorem.
Also note that any $x$-derivatives $\partial_{jA} \psi_x$,
$\partial_{jA}\partial_{kB} \psi_x$, etc. still lie in $\mathcal{S}_n$.

$\psi_x$ is $SU(N)$-invariant (covariant) in the sense that
$\tilde{R}\psi_{Rx}(Rz) = \psi_x(z)$, where
$R$ (resp. $\tilde{R}$) $\in SU(N) \hookrightarrow SO(n)$ (resp. $\Spin(\mathcal{F}_n)$).
This follows from the uniqueness of $\psi_x$ at each point $x$
and covariance of the operator $H_D(x)$, i.e.
$U_R H_D(x) U_R^\dagger = H_D(R^Tx)$,
where $U$ denotes the corresponding unitary representation
of $SU(N)$ on $\mathpzc{h}$.
\qed

\subsection{Proof of Proposition 2}

Since $\psi_x(z)$ is Gaussian in the $z$-coordinates, the evaluation of the moments
$\left\| |z|^k \psi_x \right\|_\mathpzc{h}^2 = \langle |z|^{2k} \rangle_{\psi_x}$
is straightforward. We find that
$\|\psi_x\|_{\mathpzc{h}}^2 = \|\xi_x\|_{\mathcal{F}}^2 = 1$,
\begin{eqnarray*}
    \left\| |z| \psi_x \right\|_\mathpzc{h}^2 &=&
    \sum_{A}\frac{1}{\omega_A},
\\
    \left\| |z|^2 \psi_x \right\|_\mathpzc{h}^2 &=&
    \frac{3}{2}\sum_A \frac{1}{\omega_A^2} +
    \frac{1}{2}\left( \sum_A \frac{1}{\omega_A} \right)^2, \qquad \textrm{and}
\\
    \left\| |z|^4 \psi_x \right\|_\mathpzc{h}^2 &=&
    k_1 \sum_A \frac{1}{\omega_A^4} +
    k_2 \left(\sum_A \frac{1}{\omega_A^3} \right) \left( \sum_A \frac{1}{\omega_A} \right) +
    k_3 \left( \sum_A \frac{1}{\omega_A^2} \right)^2
\\  &&
    +\ k_4 \left( \sum_A \frac{1}{\omega_A^2} \right) \left( \sum_A \frac{1}{\omega_A} \right)^2
    + k_5 \left( \sum_A \frac{1}{\omega_A} \right)^4
\end{eqnarray*}
for some combinatorial factors $k_1,\ldots,k_5$.
For example, the evaluation of $\left\| |z|^2 \Psi_x \right\|_\mathpzc{h}^2$ goes as follows
\begin{eqnarray*}
    \lefteqn{ \left\| |z|^2 \psi_x \right\|_\mathpzc{h}^2 = \langle |z|^4 \rangle_{\psi_x} }\\
    && = \pi^{-n} s^{\frac{1}{2}} \int \left( |u|^2 + |v|^2 \right)^2 e^{-u \cdot S^{1/2} u}e^{-v \cdot S^{1/2} v} d^nud^nv \\
    && = 2\pi^{-\frac{n}{2}}s^{\frac{1}{4}} \int |u|^4 e^{-u \cdot S^{1/2} u} d^nu
    \ +\ 2\pi^{-n}s^{\frac{1}{2}} \left( \int |u|^2 e^{-u \cdot S^{1/2} u} d^nu \right)^2 \\
    && = 2\pi^{-\frac{n}{2}}s^{\frac{1}{4}} \sum_{A,B} \int \tilde{u}_A^2\tilde{u}_B^2 e^{-\sum_C \omega_C \tilde{u}_C^2} d^n\tilde{u} \\
    &&\quad +\ 2\pi^{-n}s^{\frac{1}{2}} \left( \sum_A \int \tilde{u}_A^2 e^{-\sum_C \omega_C \tilde{u}_C^2} d^n\tilde{u} \right)^2 \\
    && = 2\sum_A \frac{1\cdot 3}{2^2 \omega_A^2} + 4\sum_{A<B} \frac{1}{2\omega_A} \frac{1}{2\omega_B}
    + 2\left( \sum_A \frac{1}{2 \omega_A} \right)^2\\
    && = \frac{3}{2}\sum_A \frac{1}{\omega_A^2} + \frac{1}{2}\left(\sum_A \frac{1}{\omega_A} \right)^2
    = \frac{3}{2} \tr S^{-1} + \frac{1}{2} \left( \tr S^{-\frac{1}{2}} \right)^2,
\end{eqnarray*}
where we diagonalized $S = R^T [\omega_A]^2 R$ (at the point $x$) and put $\tilde{u} := Ru$.
Hence,
\begin{eqnarray*}
    \left\| |z| \psi_x \right\|_\mathpzc{h} \le \sqrt{\frac{n}{\omega_{\min}}}, \quad
    \left\| |z|^2 \psi_x \right\|_\mathpzc{h} \le \frac{\sqrt{n(n+3)/2}}{\omega_{\min}}, \ \ \textrm{and} \ \
    \left\| |z|^4 \psi_x \right\|_\mathpzc{h} \le \frac{C_4}{\omega_{\min}^2}
\end{eqnarray*}
for some positive constant $C_4$.
\qed

\subsection{Proof of Proposition 3}

We divide the proof into two steps.
First, we show that the conditions \eqref{prop_chi_support}
can be satisfied on some subset $D_t \subseteq \mathbb{R}^{7n}$.
Then we construct a function $\chi_t$ with support on $D_t$ which
also satisfies the conditions \eqref{prop_chi_norms}.

\subsubsection{Construction of the set $D_t$}

We start by finding an explicit point $\hat{x}$ where $\omega_{\min}(\hat{x}) > 0$.
A basis for the Lie algebra $i\mathfrak{su}(N)$ of traceless hermitian
$N\times N$-matrices is given by $N-1$ diagonal ones, $h_k$,
together with the off-diagonal
$$
    e_{ij} :=  E_{ij} + E_{ji}, \qquad
    f_{ij} := i(E_{ij} - E_{ji}),
$$
($1 \le i<j \le N$), where $E_{ij}$ denotes the standard basis of matrices.
For any diagonal matrix $\Lambda = \diag(\lambda_1,\ldots,\lambda_N)$
we have
\begin{equation} \label{comm_rels}
    [\Lambda,e_{ij}] = -i(\lambda_i - \lambda_j)f_{ij} \quad \textrm{and} \quad
    [\Lambda,f_{ij}] =  i(\lambda_i - \lambda_j)e_{ij}.
\end{equation}
Let e.g. $\hat{X}_1 := \diag(m,m-1,\ldots,-m+1,-m)$
(or any other traceless diagonal matrix with all entries different),
and $\hat{X}_2 := \sum_{i<j} e_{ij}$.
Now, take any fixed
$$
    E = \sum_k \alpha_k h_k + \sum_{i<j}\beta_{ij}e_{ij} + \sum_{i<j}\gamma_{ij}f_{ij} \in i\mathfrak{su}(N)
$$
and require $E$ to commute with both $\hat{X}_1$ and $\hat{X}_2$.
Then, by \eqref{comm_rels}, $[\hat{X}_1,E]=0$ implies $\beta_{ij} = \gamma_{ij} = 0$,
i.e. $E$ must be diagonal, $E = \diag(\lambda_1,\ldots,\lambda_N)$.
Again, by \eqref{comm_rels}, $[E,\hat{X}_2]=0$ implies
$\sum_{i<j}(\lambda_i - \lambda_j)f_{ij}=0$,
i.e. all $\lambda_i$ are equal.
Tracelessness then implies that $E=0$. Hence,
$$
    \|[\hat{X}_1,E]\|^2 + \|[\hat{X}_2,E]\|^2 > 0
$$
for all $E \neq 0$, and since $S^{n-1}$ is compact it also follows that
$$
    \omega_{\min}^2(\hat{x}) \ge \min_{\|E\|=1} \left( \|[\hat{X}_1,E]\|^2 + \|[\hat{X}_2,E]\|^2 \right) =: c > 0,
$$
where $\hat{x} = (\hat{x}_1,\hat{x}_2,x_3,\ldots,x_7) \in \mathbb{R}^{7n}$,
and $\hat{x}_1 \leftrightarrow \hat{X}_1$, $\hat{x}_2 \leftrightarrow \hat{X}_2$ as usual.

Now, consider the map
$F\!: \mathbb{R}^n \times \mathbb{R}^n \times S^{n-1} \to \mathbb{R}_+$,
$$
    F(x_1,x_2,e) := \|[X_1,E]\|^2 + \|[X_2,E]\|^2.
$$
We know from the above that $F(\hat{x}_1,\hat{x}_2,\cdot) \ge c$.
Furthermore, note that for any $R \in SU(N) \hookrightarrow SO(n)$,
since $F(Rx_1,Rx_2,e) = F(x_1,x_2,R^Te)$,
we have $F(R\hat{x}_1,R\hat{x}_2,\cdot) \ge c > 0$ as well.
Then, because $F$ is continuous and $S^{n-1}$ compact, there exists an $\epsilon_R > 0$
such that $F(x_1,x_2,\cdot) \ge c/2$ for all
$x_1, x_2$ in the balls $B_{\epsilon_R}(R\hat{x}_1)$ and $B_{\epsilon_R}(R\hat{x}_2)$, respectively.
Also, by compactness of $SU(N)$, there is an $\epsilon > 0$ such that
$$
    F(x_1,x_2,\cdot) \ge \frac{c}{2}
    \quad \forall (x_1,x_2) \in B_{\epsilon}(R\hat{x}_1) \times B_{\epsilon}(R\hat{x}_2)
    \quad \forall R \in SU(N).
$$
Therefore, defining
$$
    D_1 := \bigcup_{R \in SU(N)} B_{\epsilon}(R\hat{x}_1) \times B_{\epsilon}(R\hat{x}_2)
    \times B_1(0)^5 \ \subseteq \ \mathbb{R}^{7n},
$$
we find that $\omega_{\min}(x) \ge c/2$ for all $x \in D_1$.
Furthermore, we have
$$
    (|\hat{x}_1|-\epsilon)^2 + (|\hat{x}_2|-\epsilon)^2 \le |x|^2 \le (|\hat{x}_1|+\epsilon)^2 + (|\hat{x}_2|+\epsilon)^2 + 5
$$
on $D_1$. By rescaling this set (note that $F$ is homogeneous of degree $2$),
$D_t := tD_1$, we reach the conditions \eqref{prop_chi_support}.
It is also useful to note that
$\omega_{\min} \le \omega_{\max} = \|S^{\frac{1}{2}}\|_{\op} \le c_{13}|x|$
(where $\|\cdot\|_{\op}$ denotes the operator norm).

\subsubsection{Construction of the function $\chi_t$}

We set
$$
    \chi_t(x_1,\ldots,x_7) := \mu_t(x_1,x_2) \eta_t(x_3) \ldots \eta_t(x_6) \zeta_t(x_7),
$$
where $\mu_t$, $\eta_t$ and $\zeta_t$ are to be defined below.

Given some spherically symmetric bump function
$\eta \in C_0^\infty(\mathbb{R}^n)$ with
support on the unit ball $B_1(0)$ and unit $L^2$-norm, $\|\eta\|_{\mathbb{R}^n} = 1$,
we define $\eta_t(x) := t^{-n/2}\eta(x/t)$ so that
$\supp \eta_t \subseteq B_t(0)$, $\|\eta_t\|_{\mathbb{R}^n} = 1$,
and $\|\partial^\alpha \eta_t\|_{\mathbb{R}^n} = t^{-|\alpha|} \|\partial^\alpha \eta\|_{\mathbb{R}^n}$
for any partial derivative multi-index $\alpha$.

The function $\zeta_t$ is chosen to be asymptotically a gauge
invariant solution to the Helmholtz equation, namely we take
(for the case $\lambda=0$ we instead take $\zeta_t := \eta_t$)
$$
    \zeta_t(x) := A_t \rho_t(x) h(x),
$$
where $A_t := \|\rho_t h\|_{\mathbb{R}^n}^{-1}$,
$\rho_t$ is a cut-off function $\rho_t(x) := \rho(x/t)$
such that $\rho \in C_0^\infty(\mathbb{R}^n)$ is spherically symmetric,
$0 \le \rho \le 1$, $\rho=1$ on $B_{1/2}(0)$ and $\rho=0$ outside $B_1(0)$,
and
$$
    h(x) := \frac{1}{\lambda^{\frac{1}{2}}|x|} \sin_{n-3}(\lambda^{\frac{1}{2}}|x|)
$$
satisfies $(\Delta + \lambda)h = 0$ (see \cite{maciej}), with
$$
    \sin_p(x) := \sum_{k=0}^\infty \frac{(-1)^k x^{2k+1}}{2(3+p)4(5+p) \cdots 2k(2k+1+p)}
    = c_p x^{(1-p)/2} J_{(1+p)/2}(x).
$$
Since the Bessel functions $J_k$ behave asymptotically as \cite{ABR}
$$
    J_k(x) = \sqrt{\frac{2}{\pi x}} \left( \cos\left(x - \frac{\pi}{4}(2k+1)\right) + O\!\left(\frac{1}{x}\right) \right),
$$
one finds
$$
    \int_{B_R(0)} h^2 dx
    = c_{14} \int_0^R J_{(n-2)/2}^2(\lambda^{\frac{1}{2}}r) rdr
    = c_{15}R + o(R)
$$
and
$$
    \int_{B_R(0)} |\partial_A h|^2 dx
    \le c_{16} \int_0^R J_{n/2}^2(\lambda^{\frac{1}{2}}r) rdr
    \le c_{17} + c_{18}R.
$$
Hence, $A_t \le c_{19}/t^{1/2} \to 0$, $t \to \infty$, and
\begin{eqnarray*}
    \|\partial_A \zeta_t\|_{\mathbb{R}^n}
    &\le& A_t \left( \|(\partial_A\rho_t)h\|_{\mathbb{R}^n} + \|\rho_t \partial_A h\|_{\mathbb{R}^n} \right) \\
    &\le& \frac{c_{19}}{t^{1/2}}\left( \frac{c_{20}}{t}(c_{15}t)^{1/2} + (c_{17} + c_{18}t)^{1/2} \right) \to c_{21}.
\end{eqnarray*}
Furthermore,
$
    \left(\sum_A\partial_A^2 + \lambda\right)\zeta_t = A_t(\Delta\rho_t)h + 2A_t\sum_A (\partial_A\rho_t)(\partial_A h),
$
so that
$$
    \|(-\Delta - \lambda)\zeta_t\|_{\mathbb{R}^n}
    \le \frac{c_{19}}{t^{1/2}}\left( \frac{c_{22}}{t^2}(c_{15}t)^{1/2} + 2n\frac{c_{20}}{t}(c_{17} + c_{18}t)^{1/2} \right) \to 0.
$$

Lastly, we set $\mu_t(x_1,x_2) := t^{-n} \|\mu\|_{\mathbb{R}^{2n}}^{-1} \mu(x_1/t,x_2/t)$, where
$$
    \mu(x_1,x_2) := \int_{R \in SU(N)} \eta_{\epsilon}(x_1 - R\hat{x}_1) \eta_{\epsilon}(x_2 - R\hat{x}_2)\ d\mu_H(R)
$$
and $\mu_H$ denotes some Haar measure on $SU(N)$.
Then $\mu_t \in C_0^{\infty}(\mathbb{R}^{2n})$,
$\supp \mu_t \times B_t(0)^5 \subseteq D_t$,
$\|\mu_t\|_{\mathbb{R}^{2n}} = 1$,
$\|\partial^\alpha \mu_t\|_{\mathbb{R}^{2n}} \le c_{\alpha}/t^{|\alpha|}$,
and $\mu_t(Rx_1,Rx_2) = \mu_t(x_1,x_2)$ for all $R \in SU(N)$.

Hence, $\chi_t \in C_0^{\infty}(\mathbb{R}^{7n})$ is $SU(N)$-invariant,
$\supp \chi_t \subseteq D_t$, and
$$
    \|\chi_t\|_{\mathbb{R}^{7n}}^2
    = \|\mu_t\|_{\mathbb{R}^{2n}}^2 \|\eta_t\|_{\mathbb{R}^{n}}^8 \|\zeta_t\|_{\mathbb{R}^{n}}^2 = 1.
$$
Furthermore, as $t \to \infty$,
$$
    \|\partial_{jA} \chi_t\|_{\mathbb{R}^{7n}} = \left\{
    \begin{array}{ll}
        \|\partial_{jA}\mu_t\|_{\mathbb{R}^{2n}} \le c_{23}/t \to 0, \quad & j=1,2 \\
        \|\partial_A\eta_t\|_{\mathbb{R}^{n}} \le c_{24}/t \to 0, & j=3,4,5,6 \\
        \|\partial_A\zeta_t\|_{\mathbb{R}^{n}} \le c_{21}, & j=7 \\
    \end{array}
    \right.
$$
and
$$
    \|(-\Delta_x - \lambda)\chi_t\|_{\mathbb{R}^{7n}}
    \le \|\Delta_{(x_1,x_2)}\mu_t\|_{\mathbb{R}^{2n}}
    + \sum_{j=3}^6 \|\Delta_{x_j}\eta_t\|_{\mathbb{R}^{n}}
    + \|(\Delta_{x_7} - \lambda)\zeta_t\|_{\mathbb{R}^{n}} \to 0.
$$
\qed

\subsection{Proof of Lemma 4}

Here, we will denote the partial derivatives
$\partial_{jA}\psi_x$, $\partial_{jA}^2\psi_x$ by $\psi_x'$ and $\psi_x''$, respectively.
Since $\psi_x$ is a zero-energy state of $H_D$,
we have in particular that $\partial_{jA}(H_D \psi_x) = 0$,
which can be equivalently written as
$$
    -H_D \psi_x' = \left( \bar{z} \cdot S'(x) z + 2W'(x)\lambda \lambda^{\dagger} \right)\psi_x =: \Phi_x.
$$
In the following we will need an estimate on the norm of $\Phi_x \in \mathcal{S}_n$.
We have, using Propositions \ref{prop_z_estimates} and \ref{prop_chi},
\begin{eqnarray*}
    \|\Phi_x\|_{\mathpzc{h}} &\le& \left\| \bar{z} \cdot S'(x) z \psi_x \right\|_{\mathpzc{h}}
    + 2\left\| W'(x)\lambda\lambda^{\dagger}\psi_x \right\|_{\mathpzc{h}}
\\[5pt]
    &\le& \left\|S'(x)\right\|_{\op(\mathbb{R}^n)} \left\| |z|^2 \psi_x \right\|_{\mathpzc{h}}
    + 2\left\| W'(x)\lambda\lambda^{\dagger} \right\|_{\op(\mathcal{F})}\left\| \psi_x \right\|_{\mathpzc{h}}
\\
    &\le& c_{25}|x|\frac{C_2}{\omega_{\min}(x)} + c_{26} \le c_{27}\frac{t}{t} + c_{26} =: c_5.
\end{eqnarray*}
Now, we note that $\Phi_x \in (\ker \bar{H}_D)^{\bot}=:\mathcal{P}_+$ where $\bar{H}_D$ denotes the
self-adjoint extension of $H_D$.
To see this, consider any $\Phi_0 \in \ker \bar{H}_D$. We have
$$
    \langle \Phi_x,\Phi_0 \rangle_{\mathpzc{h}}
    = -\langle \bar{H}_D\psi_x',\Phi_0 \rangle_{\mathpzc{h}}
    = -\langle \psi_x',\underbrace{\bar{H}_D\Phi_0}_{=0} \rangle_{\mathpzc{h}} = 0.
$$
Therefore,
since the lowest eigenvalue of $\bar{H}_D$ on $\mathcal{P}_+$ is $\omega_{\min} > 0$, we have
$\left\|\bar{H}_D|_{\mathcal{P}_+}^{-1} \right\|_{\op(\mathpzc{h})} \le \frac{1}{\omega_{\min}}$
and
$$
    \left\| \psi_x' \right\|_{\mathpzc{h}}
    = \left\| \bar{H}_D^{-1}\Phi_x \right\|_{\mathpzc{h}}
    \le \left\| \bar{H}_D|_{\mathcal{P}_+}^{-1} \right\|_{\op(\mathpzc{h})} \left\| \Phi_x \right\|_{\mathpzc{h}}
    \le \frac{c_5}{\omega_{\min}}.
$$

Similarly, taking the second derivative we have
$\partial_{jA}^2(H_D \psi_x) = 0$, i.e.
$$
    -H_D \psi_x'' = \bar{z} \cdot S''(x) z \psi_x
    + 2\left( \bar{z} \cdot S'(x) z + 2W'(x)\lambda \lambda^{\dagger} \right) \psi_x'
    =: \tilde{\Phi}_x.
$$
Just as above, we see that $\tilde{\Phi}_x \in \mathcal{P}_+ \cap \mathcal{S}_n$,
but we will also need an estimate on $\| |z|^2 \psi_x' \|_{\mathpzc{h}}$.
For this, we recall from the proof of Proposition \ref{prop_z_estimates} that
$$
    \langle |z|^2 \psi_x, |z|^2 \psi_x \rangle_{\mathpzc{h}}
    = \frac{3}{2} \tr S^{-1} + \frac{1}{2} \left( \tr S^{-\frac{1}{2}} \right)^2 =: T(x),
$$
Note that $T$ is smooth for $s > 0$ and homogeneous of degree $-2$,
because $S$ is homogeneous of degree $2$.
It follows that, if $x = re$ with $e \in S^{7n-1}$, then
$T''(x) = r^{-4}T''(e)$ and
\begin{eqnarray*}
    2\re\ \langle |z|^2 \psi_x'', |z|^2 \psi_x \rangle_{\mathpzc{h}}
    \ +\ 2\left\| |z|^2 \psi_x' \right\|_{\mathpzc{h}}^2
    \ =\ T''(x) \le \frac{1}{r^4} \sup_{K_0} |T''|,
\end{eqnarray*}
where $K_0 := S^{7n-1} \cap \{x \in \mathbb{R}^{7n} : s(x) \ge (c_1/c_3)^{2n}\}$
is compact and we have used that $\omega_{\min}/|x| \ge c_1/c_3$ by Proposition \ref{prop_chi}.
Hence,
\begin{eqnarray*}
    \left\| |z|^2 \psi_x' \right\|_{\mathpzc{h}}^2
    \le \frac{c_{28}}{|x|^4} + \left\| \psi_x'' \right\|_{\mathpzc{h}} \left\| |z|^4 \psi_x \right\|_{\mathpzc{h}}
    \le \frac{c_{29}}{\omega_{\min}^4} \left( 1 + \omega_{\min}^2 \|\psi_x''\|_{\mathpzc{h}} \right),
\end{eqnarray*}
so that
\begin{eqnarray*}
    \lefteqn{ \|\psi_x''\|_{\mathpzc{h}}
    = \| H_D^{-1} \tilde{\Phi}_x \|_{\mathpzc{h}} }\\
&&  \le \frac{1}{\omega_{\min}} \left(
        c_{30} \left\| |z|^2 \psi_x \right\|_{\mathpzc{h}}
        + c_{31} \|S'\|_{\op} \left\| |z|^2 \psi_x' \right\|_{\mathpzc{h}}
        + c_{32} \|\psi_x'\|_{\mathpzc{h}}
    \right) \\
&&  \le \frac{1}{\omega_{\min}} \left(
        \frac{c_{30}C_2}{\omega_{\min}}
        + \frac{c_{33}|x|}{\omega_{\min}^2} \left( 1 + \omega_{\min}^2 \|\psi_x''\|_{\mathpzc{h}} \right)^{\frac{1}{2}}
        + \frac{c_{32}c_5}{\omega_{\min}}
    \right)
\end{eqnarray*}
and thus, $\omega_{\min}^2 \|\psi_x''\|_{\mathpzc{h}} \le c_6$
for some constant $c_6$.
\qed

\section{Summary}

We have introduced $G_2 \times U(1) \times SU(N)$ invariant
matrix models as deformations of the standard $\Spin(9) \times SU(N)$ invariant
models by conjugating a supercharge with a cubic, octonionic, exponential.
Furthermore, similarly to what has been shown for the original models,
we have proved that the spectrum of the corresponding Hamiltonian $\tilde{H}$
covers the whole positive half-axis
by finding sequences of states contradicting existence of a bounded inverse
to the operator $\tilde{H}-\lambda$ for any $\lambda \ge 0$.
However, contrary to the case for the original models, we have constructed such sequences
explicitly, \emph{without} fixing the gauge.
Making use of the convenient structure of terms appearing in
$\tilde{H}$, we could configure the states to annihilate some terms,
while, related to having the possibility of making the lowest
eigenvalue of a certain frequency matrix $S$ arbitrarily large,
other terms could be made arbitrarily small -- using a gauge invariant
asymptotic solution to the Helmholtz equation, with support on a set
of matrices that are not simultaneously block-diagonalizable.

\vspace{12pt}
\noindent{\bf Acknowledgements}

This work was supported by the Swedish Research Council and the
Marie Curie Training Network ENIGMA (contract MRNT-CT-2004-5652).

J.H. and D.L. would like to thank V. Bach for collaboration on related subjects.
D.L. would also like to thank H. Kalf, L. Svensson and M. Bj\"orklund for discussions.

\newpage

\section*{Appendix A }

In this appendix we give notation and conventions used in the paper (cp. e.g. \cite{Hopperev}).

The supermembrane matrix theory is a quantum mechanical model with
$\mathcal{N}=16$ supersymmetries, $SU(N)$ gauge invariance and
$\Spin(9)$ symmetry. The theory involves real bosonic variables
$x_{sA}$ (coordinates) and real fermionic ones $\theta_{\alpha A}$
(Majorana spinors)  with $s=1,\ldots,9$, $\alpha=1,\ldots16$ and
$A=1,\ldots,N^2-1$ - spatial, spinor and color indices respectively.
The corresponding supercharges and the Hamiltonian of the model are
$$
    \mathcal{Q}_{\alpha} = \Big( p_{sA}\gamma^s_{ \alpha \beta}
    + \frac{1}{2}f_{ABC}x_{sB}x_{tC}\gamma^{st}_{ \alpha \beta} \Big)\theta_{\beta A}, \quad
    \gamma^{st}=\frac{1}{2}[\gamma^s,\gamma^t],
$$
$$
    H = p_{sA}p_{sA} + \frac{1}{2}(f_{ABC}x_{sB}x_{tC})^2
    + if_{ABC}\gamma^s_{\alpha \beta} \theta_{\alpha A}\theta_{\beta B} x_{sC},
$$
\begin{equation}
    \{\mathcal{Q}_{\alpha}, \mathcal{Q}_{\beta}\}
    = \delta_{\alpha\beta}H+2\gamma^s_{\alpha\beta}x_{sA}J_A, \quad
    J_A = f_{ABC}\Big( x_{sB}p_{sC} - \frac{i}{2}\theta_{\alpha B}\theta_{\alpha C} \Big).
    \label{symqm}
\end{equation}
Here $p_{sA}$ are momenta conjugate to $x_{sA}$,
$[x_{sA},p_{tB}]=i\delta_{st}\delta_{AB}$, $\gamma^s$ are
$16 \times 16$ dimensional, real matrices s.t.
$\{\gamma^s,\gamma^t\}=2\delta^{st}\textbf{1}_{16\times 16}$,
$\theta_{\alpha A}$ are Grassmann numbers s.t.
$\{\theta_{\alpha A},\theta_{\beta B} \}=\delta_{\alpha \beta}\delta_{AB}$,
and $f_{ABC}$ are $SU(N)$ structure constants (real, antisymmetric).
The operators are defined on the Hilbert space
$\mathcal{H}=L^2(\mathbb{R}^{9(N^2-1)})\otimes \mathcal{F}$, where
$\mathcal{F}$ is the irreducible representation of $\theta$'s, while
the physical (gauge invariant) Hilbert space consists of states $|
\psi \rangle $ satisfying $J_A| \psi \rangle =0$ which corresponds
to the Gauss law in unreduced $\mathcal{N}=1$ super Yang-Mills theory.

Such singlet constraint is an essential requirement for the model to
be supersymmetric which is apparent in Eq. \eqref{symqm}. However,
the necessity of the constraint follows also from simply counting
the fermionic and bosonic degrees of freedom. Let us consider the
Fock space formulation of the model.
For the case at hand
there are $9(N^2-1)$ bosonic degrees of freedom, however there
are $\frac{16}{2}(N^2-1)$ fermionic ones. The mismatch is equal to
$N^2-1$, which is exactly the number of constraints coming from the
Gauss law.

There are many ways in which one can single out 8 out of 16 fermions
(which is required in order to obtain an irreducible Fock representation $\mathcal{F}$).
We will follow the convention in \cite{dWHN} and introduce complex spinor variables
$\lambda_{\alpha A} := \frac{1}{\sqrt{2}}(\theta_{\alpha A} + i\theta_{8+\alpha \thinspace A})$
i.e.\footnote{Other choices of 8 fermions are possible, e.g. Majorana-Weyl spinors
(see \cite{Wosiek}). \\
From now on the spinor indices $\alpha, \beta, \ldots$ run from 1 to 8.  }
$$
    \theta_{\alpha A}=\frac{1}{\sqrt{2}}(\lambda_{\alpha A}+\lambda_{\alpha A}^{\dagger}),
    \quad \theta_{\alpha+8 \thinspace A}=\frac{1}{i\sqrt{2}}(\lambda_{\alpha A}-\lambda_{\alpha A}^{\dagger}).
$$
We then also split the coordinates $x_{sA}$
into $(x_{jA},z_A,\bar{z}_A)$
where $z_A=x_{8A}+ix_{9A}$ and $j=1,\ldots,7$.

After this is done the $\Spin(9)$ symmetry of \eqref{symqm} is not
explicit, however now an arbitrary wavefunction $\Psi(x,z,\bar{z})$
can be written as
$$
    \Psi(x,z,\bar{z}) = \psi + \psi_{\alpha A}\lambda_{\alpha A}
    + \frac{1}{2!}\psi_{\alpha A \thinspace \beta   B}\lambda_{\alpha A}\lambda_{\beta B}+\ldots \ ,
$$
with $\psi_{\alpha_1 A_1 \ldots \alpha_lA_l}$
complex-valued and square integrable.
The above sum is finite and truncates when the number of fermions is
more than $8(N^2-1)$.\footnote{Note that in this notation
$\lambda_{\alpha A}$ is a fermionic creation operator while
$\lambda_{\alpha A}^{\dagger}$ fermionic annihilation operator.}

It now follows that the Hamiltonian \eqref{symqm} can be written in
terms of non-hermitian (``cohomology'') charges
$Q_{\alpha} := \frac{1}{\sqrt{2}}(\mathcal{Q}_{\alpha} + i\mathcal{Q}_{8+\alpha})$,
$$
    Q_{\beta}=\left(i\delta_{\alpha
    \beta}\frac{i}{2}f_{ABC}z_B\bar{z}_C +  i\Gamma_{\alpha \beta}^j
    \frac{\partial}{\partial
    x_{jA}}-\frac{1}{2}f_{ABC}x_{jB}x_{kC}\Gamma^{jk}_{\alpha\beta}
    \right)\lambda_{\alpha A}
$$
\begin{equation}
    + \left(2\delta_{\alpha\beta}\frac{\partial}{\partial z_A}
    -if_{ABC}x_{jB}\bar{z}_C \Gamma^j_{\alpha\beta}
    \right) \lambda_{\alpha A}^{\dagger},
\end{equation}
so that, on the physical Hilbert space,
$$
    \{Q_{\alpha},Q_{\beta}^{\dagger}\}=\delta_{\alpha \beta} H,  \ \ \ \
    \{Q_{\alpha},Q_{\beta}\}=0, \ \ \ \
    \{Q_{\alpha}^{\dagger},Q_{\beta}^{\dagger}\}=0.
$$
Here, $\Gamma^j$ are $8 \times 8$, purely imaginary, antisymmetric matrices
satisfying $\{\Gamma^j,\Gamma^k\}=2\delta^{jk}\mathbf{1}_{8 \times 8}$.
We have chosen the following representation of $\gamma^s$ matrices
$$
    \gamma^j=\left[ \begin{array}{cc}
        0 & i\Gamma^j     \\
        -i\Gamma^j & 0   \\
    \end{array} \right], \quad
    \gamma^8=\left[ \begin{array}{cc}
        0 & \textbf{1}_{8 \times 8}       \\
        \textbf{1}_{8 \times 8} & 0   \\
    \end{array} \right], \quad
    \gamma^9=\left[ \begin{array}{cc}
        \textbf{1}_{8 \times 8} & 0       \\
        0 & -\textbf{1}_{8 \times 8}   \\
    \end{array} \right],
$$
implying
$$
    \gamma^{jk}=\left[ \begin{smallmatrix}
        \Gamma^{jk} & 0  \\[3pt]
        0 & \Gamma^{jk}
    \end{smallmatrix} \right], \
    \gamma^{j8}=\left[ \begin{smallmatrix}
        i\Gamma^{j} & 0  \\[3pt]
        0 & -i\Gamma^{j}
    \end{smallmatrix} \right], \
    \gamma^{j9}=\left[ \begin{smallmatrix}
        0 &  -i\Gamma^{j}  \\[3pt]
        -i\Gamma^{j}  & 0
    \end{smallmatrix} \right], \
    \gamma^{89}=\left[ \begin{smallmatrix}
        0 &  -\textbf{1}_{8 \times 8}  \\[3pt]
        \textbf{1}_{8 \times 8}   & 0
    \end{smallmatrix} \right],
$$
and
$$
    \gamma^{jkl}=\left[ \begin{smallmatrix}
        0 &   i\Gamma^{jkl}  \\[3pt]
        -i\Gamma^{jkl} & 0
    \end{smallmatrix} \right], \
    \gamma^{jk8}=\left[ \begin{smallmatrix}
        0 &   \Gamma^{jk}  \\[3pt]
        \Gamma^{jk} & 0
    \end{smallmatrix} \right], \
    \gamma^{jk9}=\left[ \begin{smallmatrix}
        \Gamma^{jk}   &  0  \\[3pt]
        0 & -\Gamma^{jk}
    \end{smallmatrix} \right], \
    \gamma^{j89}=\left[ \begin{smallmatrix}
        i\Gamma^{j}   &  0  \\[3pt]
        0 & i\Gamma^{jk}
    \end{smallmatrix} \right],
$$
where $\gamma^{st}:=\frac{1}{2}[\gamma^s,\gamma^t]$,
$\gamma^{stu}:=\frac{1}{6}(\gamma^s[\gamma^t,\gamma^u] + \textrm{cycl.})$ and
$\Gamma^{jk}$, $\Gamma^{jkl}$ respectively.

It is here where the octonions enter, in
choosing the representation $i\Gamma^j_{\alpha 8}=\delta^j_{\alpha}$,
$i\Gamma^j_{kl}=-c_{jkl}$ with totally antisymmetric octonionic
structure constants.\footnote{Explicitly, $c_{ijk}=+1$ for
$(ijk)=(123),(165),(246),(435),(147),(367),(257)$.}
This is also natural from the view of representation theory of Clifford algebras
since the representations of $\Gamma^j$ are uniquely
given by left or right multiplication on the octonion algebra
(see e.g. \cite{Lawson-Michelsohn}).
Furthermore, because the automorphism group of the octonions is given
by the exceptional group $G_2$ 
(which is also the subgroup of $\Spin(7)$ fixing a chosen spinor index),
the deformed Hamiltonians $H_k$, $\hat{H}$, and $\tilde{H}$ will be $G_2$ invariant.

\section*{Appendix B }

Starting from the 9-dimensional Fierz identity (see e.g. \cite{baake})
$$
    \gamma^s_{\alpha \beta } \gamma^{st}_{\alpha'\beta'} +
    \gamma^s_{\alpha'\beta } \gamma^{st}_{\alpha \beta'} +
    \gamma^s_{\alpha \beta'} \gamma^{st}_{\alpha'\beta } +
    \gamma^s_{\alpha'\beta'} \gamma^{st}_{\alpha \beta }
    = 2(\delta_{\alpha\alpha'} \gamma^t_{\beta\beta'} - \delta_{\beta\beta'} \gamma^t_{\alpha\alpha'}),
$$
which holds for all $t = 1,\ldots,9$, $\alpha,\alpha',\beta,\beta' =
1,\ldots,16$, and using the representation in Appendix A with
$\alpha,\alpha',\beta' = 1,\ldots,8$, $\beta = 9,\ldots,16$ (then
redefining $\beta := \beta - 8$), we obtain the corresponding
7-dimensional Fierz identity
$$
    \Gamma^j_{\alpha \beta } \Gamma^{jk}_{\alpha'\beta'} +
    \Gamma^j_{\alpha'\beta } \Gamma^{jk}_{\alpha \beta'}
    = \delta_{\alpha \beta } \Gamma^k_{\alpha'\beta'}
    + \delta_{\alpha'\beta } \Gamma^k_{\alpha \beta'}
    - \delta_{\alpha \beta'} \Gamma^k_{\alpha'\beta }
    - \delta_{\alpha'\beta'} \Gamma^k_{\alpha \beta }
    -2\delta_{\alpha \alpha'} \Gamma^k_{\beta \beta'}
$$
for all $k = 1,\ldots,7$, $\alpha,\alpha',\beta,\beta' = 1,\ldots,8$.
From this identity it follows that
$$
    \Gamma^j_{\alpha \beta } \Gamma^{jk}_{\alpha'\beta'} -
    \Gamma^j_{\alpha'\beta'} \Gamma^{jk}_{\alpha \beta }
    = -2(\delta_{\alpha \alpha'} \Gamma^k_{\beta \beta'}
    + \delta_{\beta \beta'} \Gamma^k_{\alpha \alpha'}
    - \delta_{\alpha'\beta } \Gamma^k_{\alpha \beta'}
    + \delta_{\alpha \beta'} \Gamma^k_{\alpha'\beta } ).
$$
Multiplying this equation with $\Gamma^l_{\beta'\dot{\beta}}$,
summing over $\beta'$, and taking $\alpha' = \beta = \dot{\beta}$
to be fixed, we obtain
$$
    \Gamma^j_{\alpha\dot{\beta}} \Gamma^{jkl}_{\dot{\beta}\dot{\beta}}
    = \Gamma^{kl}_{\alpha\dot{\beta}}.
$$

\section*{Appendix C }

In this appendix we consider deformed Hamiltonians from a more general
viewpoint and show how one could be led to the particular deformation
considered in this paper.

Let us consider the algebra of $\mathcal{N}>1$ supersymmetric quantum mechanics, \\
$\{\mathcal{Q}_{\alpha},\mathcal{Q}_{\beta}\}=
\delta_{\alpha\beta}H$, and the corresponding cohomology
supercharges \footnote{We consider the $(\alpha\beta)$ as distinct pairs
of indices with $\alpha$ and $\beta$ in disjoint subsets of the
index set. This is a common construction of non-hermitian charges
involving the complex structure $i$, but other variations are
possible; see e.g. \cite{Lundholm} and references therein.}
$$
    Q_{\alpha \beta}:= \frac{1}{\sqrt{2}}(\mathcal{Q}_{\alpha} + i\mathcal{Q}_{\beta}), \quad
    Q_{\alpha \beta}^{\dagger}= \frac{1}{\sqrt{2}}(\mathcal{Q}_{\alpha} - i\mathcal{Q}_{\beta}).
$$
We have
$$
    \{Q_{\alpha \beta},Q_{\mu \nu}\}=0, \quad
    \{Q_{\alpha \beta},Q^{\dagger}_{\mu \nu}\} = \delta_{(\alpha\beta)(\mu\nu)}H.
$$
The deformed Hamiltonian
$H_{\alpha \beta}(k) := \{ Q_{\alpha\beta}(k), Q_{\alpha \beta}^{\dagger}(k) \}$
(no sum over $\alpha$,$\beta$) given by deformed cohomology supercharges
$Q_{\alpha\beta}(k) := e^{kg(x)}Q_{\alpha\beta}e^{-kg(x)}$, where
$k \in \mathbb{R}$, and $g(x)$ is some operator s.t.
$[\mathcal{Q}_{\beta},g(x)]$ commutes with $g(x)$, becomes
$$
    H_{\alpha \beta}(k) = H
    - 2ik\{\mathcal{Q}_{\alpha},[\mathcal{Q}_{\beta},g(x)]\}
    - 2k^2[\mathcal{Q}_{\beta},g(x)]^2.
$$
Substituting the supercharges \eqref{symqm} for the particular model
considered here, we obtain
\begin{eqnarray*}
    H_{\alpha \beta}(k) &=& H + k^2 (\partial_{sA}g(x))^2
    + k \gamma^{st}_{\alpha \beta} (\partial_{sA}g(x) p_{tA} - \partial_{tA}g(x) p_{sA}) \\[5pt]
&&  -\ k (\gamma^{st}\gamma^u)_{\alpha\beta} f_{ABC}x_{sA}x_{tB}\partial_{uC}g(x) \\[5pt]
&&  +\ 2ik\partial_{sA}\partial_{tB}g(x)\gamma^s_{\alpha\alpha'}\gamma^t_{\beta \beta'}\theta_{\alpha'A} \theta_{\beta'B}.
\end{eqnarray*}
Now, say we are interested in a particular deformation where
$g(x)$ is cubic in $x$ (so that $(\partial g)^2$ is quartic).
Because $\gamma^{stu}$ is totally antisymmetric, a natural choice is
$$
    g(x) = \frac{1}{6}f_{ABC}x_{sA}x_{tB}x_{uC}\gamma^{stu}_{\alpha\beta},
$$
with $\alpha < \beta$.
Taking e.g. $(\alpha,\beta)=(8,16)$ and choosing the representation
of $\gamma^s$ matrices as in Appendix A, we find that
$$
    g(x) = \frac{1}{6}f_{ABC}x_{jA}x_{kB}x_{lC}i\Gamma^{jkl}_{8,8}
    = \frac{1}{6}c_{jkl}f_{ABC}x_{jA}x_{kB}x_{lC},
$$
and that $H_{8,16}(k)$ becomes precisely $H_k$ in \eqref{H_k}.

\end{document}